\documentclass[referee]{aa}
\usepackage{graphicx}
\usepackage{txfonts}

\usepackage{algorithm}
\usepackage{algorithmic}
\usepackage{subfig}
\usepackage{float}
\usepackage{comment}
\usepackage{caption}
\usepackage{bm}
\usepackage{amsfonts}
\usepackage{amssymb}
\usepackage{array}
\usepackage{natbib}
\usepackage{tikz}
\usepackage{booktabs}

\usepackage{url}

\usepackage{color}
\usepackage{multirow}

\usepackage{multicol}
\usepackage{colortbl, dcolumn}
\usepackage{xcolor}

\newcommand{\R}{\mathbb{R}}

\begin{document}

\title{Unbiased CLEAN for STIX in Solar Orbiter}

\author{Emma Perracchione\inst{1}  \and Fabiana Camattari\inst{1,2} \and Anna Volpara\inst{3} \and Paolo Massa\inst{4} \and Anna Maria Massone\inst{3} \and Michele Piana\inst{3,5}}
\institute{Dipartimento di Scienze Matematiche \lq \lq Giuseppe Luigi Lagrange\rq \rq, Politecnico di Torino, Corso Duca degli Abruzzi, 24, 10129, Torino, Italy
 \and Dipartimento di Matematica \lq \lq Giuseppe Peano\rq \rq, Università di Torino, Via Carlo Aberto 10, 10123, Torino, Italy
\and MIDA, Dipartimento di Matematica, Università di Genova, via Dodecaneso 35 16145 Genova, Italy
\and Department of Physics and Astronomy, Western Kentucky University, Bowling Green, KY 42101, USA
\and Istituto Nazionale di Astrofisica, Osservatorio Astrofisico di Torino, Pino Torinese, Italy
\\
\email{emma.perracchione@polito.it}, \\
\email{fabiana.camattari@polito.it}, \\
\email{volpara@dima.unige.it}, \\
\email{paolo.massa@wku.edu}, \\
\email{massone@dima.unige.it}, \\
\email{piana@dima.unige.it}
}

%\date{Received September 15, 1996; accepted March 16, 1997}

\abstract
% context heading (optional), leave it empty if necessary
{}
{To formulate, implement, and validate a user-independent release of CLEAN for Fourier-based image reconstruction of hard X-rays flaring sources.}
{CLEAN is an iterative deconvolution method for radio and hard X-ray solar imaging. In a specific step of its pipeline, CLEAN requires the convolution between an idealized version of the instrumental Point Spread Function (PSF), and a map collecting point sources located at positions on the solar disk from where most of the flaring radiation is emitted. This convolution step has highly heuristic motivations and the shape of the idealized PSF, which depends on the user's choice, impacts the shape of the overall reconstruction. Here we propose the use of an interpolation/extrapolation process to avoid this user-dependent step, and to realize a completely unbiased version of CLEAN.}
{Applications to observations recorded by the Spectrometer/Telescope for Imaging X-rays (STIX) on-board Solar Orbiter show that this unbiased release of CLEAN outperforms the standard version of the algorithm in terms of both automation and reconstruction reliability, with reconstructions whose accuracy is in line with the one offered by other imaging methods developed in the STIX framework.}
{This unbiased version of CLEAN proposes a feasible solution to a well-known open issue concerning CLEAN, i.e., its low degree of automation. Further, this study provided the first application of an interpolation/extrapolation approach to image reconstruction from STIX experimental visibilities.}

\keywords{Sun: flares -- Sun: X-rays, gamma-rays -- Techniques: image processing -- Methods: data analysis -- }

\titlerunning{Unbiased CLEAN}
\authorrunning{Perracchione et al}

\maketitle

%\date{\today}
\section{Introduction}
\label{intro}

Native measurements in both radio astronomy \citep{richard2017interferometry} and modern hard X-ray solar imaging \citep{piana2022hard} are sets of spatial Fourier components of the incoming source flux, named visibilities, measured at specific spatial frequency samples, named $({\mbox{u}},{\mbox{v}})$ points. In both radio astronomy and hard X-ray solar imaging, CLEAN \citep{1974A&AS...15..417H,Schmahl} is a non-linear image reconstruction algorithm that iteratively deconvolves the instrumental point spread function (PSF) from the so-called dirty map, i.e the discretized inverse Fourier transform of the experimental visibility set. More specifically, the CLEAN algorithm is made of a CLEAN loop, which generates: a set of CLEAN components located at the points of the solar disk from where most of the source emission propagates; an estimate of the background; the convolution of the CLEAN components map with an idealized PSF, named the CLEAN beam; and, eventually, the CLEANed map, i.e., the sum of the outcome of this convolution step with the convolved background residuals.

In the framework of the visibility-based NASA Reuven Ramaty High Energy Solar Spectroscopic Imager (RHESSI) \citep{enlighten1658}, CLEAN has been by far the most utilized image reconstruction method. However, we are now in the Solar Orbiter era and other image reconstruction methods besides CLEAN are currently used for the analysis of the hard X-ray visibilities recorded by the
Spectrometer/Telescope for Imaging X-rays (STIX) on-board the ESA cluster \citep{krucker2020spectrometer}. This is due to the fact that, meanwhile, novel imaging algorithms have been introduced \citep{massa2020mem_ge,2022A&A...668A.145V,2021ApJ...919..133P,2020OAst...29..220S,2018A&A...615A..59D,2017ApJ...849...10F}, which are characterized by notable reliability and by an automation degree higher than the one offered by CLEAN. In fact, the step of the iterative scheme that requires the convolution of the CLEAN components map with the CLEAN beam is significantly dependent of the user's choice, since the functional shape of the CLEAN beam (e.g., its Full Width at Half Maximum, FWHM) is typically designed according to heuristic motivations. This results in CLEANed maps of the same event that are often characterized by different properties, while conservative choices of the CLEAN beam's FWHM often leads to under-resolved reconstructions with correspondingly high $\chi^2$ values.

The objective of the present paper is to introduce a completely user-independent technique for the exploitation of the CLEAN components associated to the analysis of STIX visibilities. In particular, we show here that the feature augmentation process introduced by \cite{2021InvPr..37j5001P} and applied to RHESSI experimental and STIX synthetic visibilities can lead to an unbiased version of CLEAN (u-CLEAN), in which the convolution between the CLEAN components map and the idealized PSF is replaced by an automated interpolation/extrapolation procedure. Further, u-CLEAN does not need any addition of residuals, since the resulting reconstructed map is automatically embedded in the emission background. In this study we show how the u-CLEAN pipeline works in the case of STIX observations and compare its outcomes with
the reconstructions provided by other imaging methods contained in the STIX ground software. We point out that the u-CLEAN reconstructions presented in this paper can be interpreted also as the first maps provided by the augmented uv$\_$smooth algorithm introduced by \cite{2021ApJ...919..133P} in the case of experimental STIX visibilities.

The plan of the paper is as follows. Section \ref{preliminari} sets up the formalism of CLEAN and points out the need of an unbiased release of the iterative algorithm. Section \ref{uclean} introduces the novel automated version of CLEAN exploiting feature augmentation. Section \ref{esperimenti} applies u-CLEAN to STIX observations and assesses its performances. Our conclusions are offered in Section \ref{conclusioni}.

\section{Toward unbiased CLEAN}
\label{preliminari}

\begin{figure}
    \centering
    \includegraphics[scale=0.2]{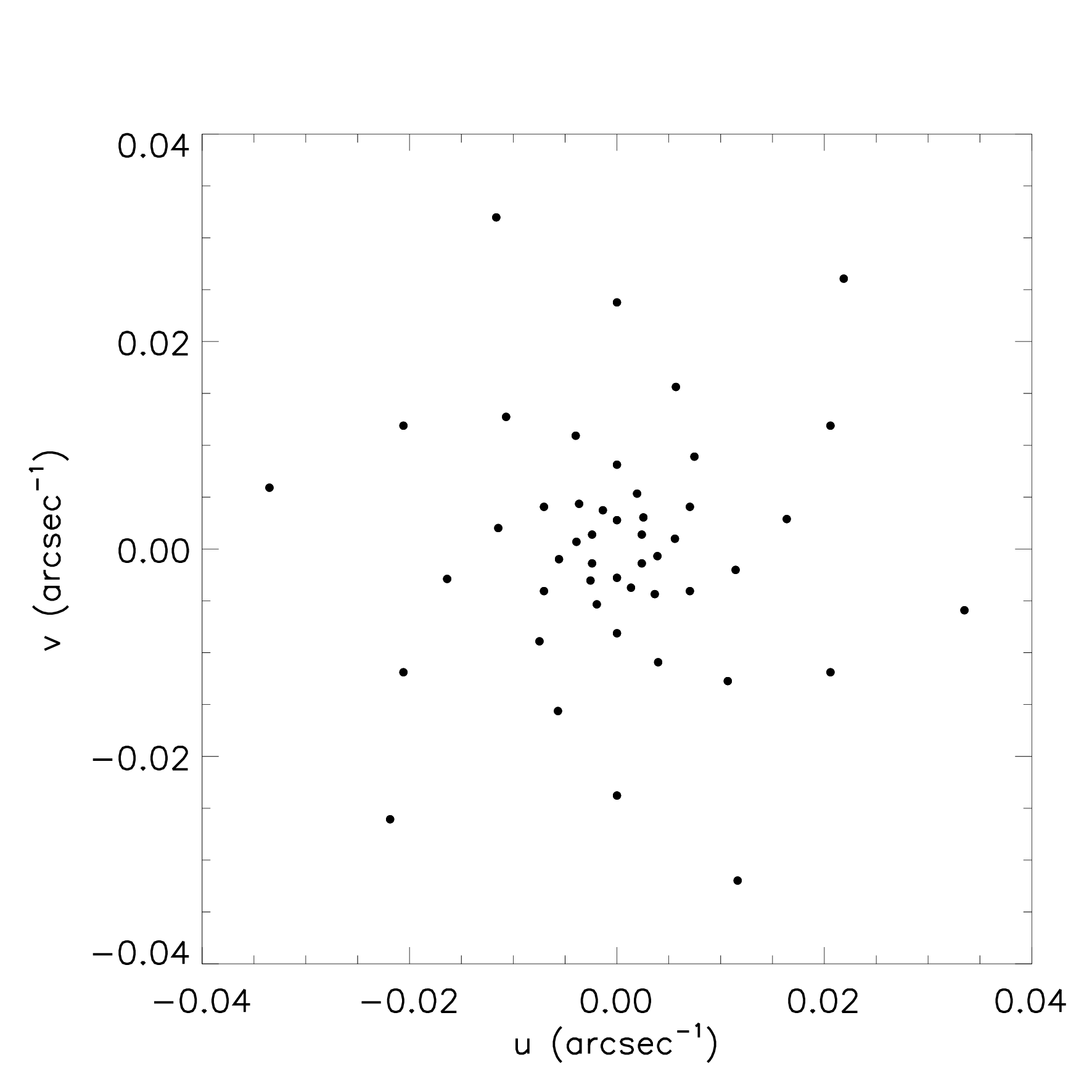}
    \caption{Sampling of the spatial frequency $({\mbox{u}},{\mbox{v}})$-plane realized by the 30 STIX collimators, which have been duplicated by exploiting the properties of the Fourier transform.}
    \label{fig:STIX-sampling}
\end{figure}

The STIX imaging concept \citep{2023arXiv230302485M} relies on Moir\'e patterns \citep{1988SoPh..118..269P} generated by 30 sub-collimators that detect photons from the Sun in the energy range between a few keV and around 100 keV. These raw data are then transformed into a set of $n=60$ visibilities that sample the spatial frequency domain, the $({\mbox{u}},{\mbox{v}})$-plane, according to the spirals in Figure \ref{fig:STIX-sampling}. Since visibilities can be seen as spatial Fourier components of the incoming photon flux, the STIX imaging problem reads as
\begin{equation}\label{eq0}
{\boldsymbol{V}} = {\boldsymbol{F}} {\boldsymbol{f}}~,
\end{equation}
where ${\boldsymbol{f}}$ is the vector whose components are the discretized values of the incoming flux, ${\boldsymbol{F}}$ is the discretized Fourier transform sampled at the set $ \{ {\boldsymbol{u}}_k=(u_k,v_k) \}_{k=1}^{n}$, and ${\boldsymbol{V}}$ is the complex vector of the observed visibilities. 

Given $\boldsymbol{x}=(x,y)$ in the image domain, visibility-based CLEAN iteratively solves  the convolution equation
\begin{equation}\label{prob}
{\hat{{\boldsymbol{f}}}}(\boldsymbol{x}) = (K * {{{\boldsymbol{f}}}})(\boldsymbol{x}) := \int\int K(\boldsymbol{x}-\boldsymbol{x}^{'}){{{\boldsymbol{f}}}}(\boldsymbol{x}^{'})d\boldsymbol{x}^{'}~,
\end{equation}
where ${{{\boldsymbol{f}}}}$ is the unknown source flux, $K$ is the so-called dirty beam, i.e. the instrumental PSF
\begin{equation}\label{dirty-beam}
K(\boldsymbol{x}) =  \sum_{k=1}^{n}  \exp(-2 \pi i \boldsymbol{x}\cdot \boldsymbol{u}_k) \delta \boldsymbol{u}_k~ ~,
\end{equation}
and ${\hat{{\boldsymbol{f}}}}$ is the so-called dirty map
\begin{equation}\label{dirty-1}
{\hat{{\boldsymbol{f}}}}(\boldsymbol{x}) =  \sum_{k=1}^{n} \boldsymbol{V}_k \exp[-2 \pi i (\boldsymbol{x}\cdot \boldsymbol{u}_k)]  \delta \boldsymbol{u}_k ,
\end{equation}
i.e., the inverse Fourier transform of the visibilities. In both equation (\ref{dirty-beam}) and equation (\ref{dirty-1})  $\delta \boldsymbol{u}_k = (\delta u_k,\delta v_k)$, $k=1,\ldots,n$, denote the weights in the numerical integration.

 CLEAN steps are summarized in a schematic way in Algorithm \ref{alg:alg1}. \\
\begin{algorithm}
\small
\caption{\textbf{CLEAN steps}}
    \hspace*{\algorithmicindent} {\textbf{Inputs:}  Sampling points $ \{ \boldsymbol{u}_k=(u_k,v_k) \}_{k=1}^{n}$; visibility vector $\boldsymbol{V}$; {gain factor} $\gamma$; CLEAN beam $K^C$. \\
    \hspace*{\algorithmicindent} \textbf{Outputs:} The CLEANed map ${\overline{{\boldsymbol{f}}}}$, the CLEAN components map ${\tilde{{\boldsymbol{f}}}}$, the convolved background residual map \hspace*{\algorithmicindent} \hspace*{\algorithmicindent}  $K^C * {{{\boldsymbol{f}}^B}}$, the dirty map ${\hat{{\boldsymbol{f}}}}$. }
\begin{algorithmic}[1]
\STATE {\bf Initialization.} \\
\begin{enumerate}
    \item[a.] {\bf{Dirty map:}} 
    \begin{equation}{\hat{{\boldsymbol{f}}}^{(0)}}(\boldsymbol{x}) =  \sum_{k=1}^{n} \boldsymbol{V}_k \exp[-2 \pi i (\boldsymbol{x}\cdot \boldsymbol{u}_k)]  \delta \boldsymbol{u}_k  .
    \end{equation}
    \item[b.] {\bf{Dirty beam:}}
    \begin{equation}
K(\boldsymbol{x}) =  \sum_{k=1}^{n}  \exp(-2 \pi i \boldsymbol{x}\cdot \boldsymbol{u}_k) \delta \boldsymbol{u}_k~ ~.
\end{equation}
\item[c.] {\bf{CLEAN component map:}}
\begin{equation}
    {\tilde{{\boldsymbol{f}}}}^{(0)} (\boldsymbol{x}) = {\boldsymbol{0}}.
\end{equation}
\end{enumerate}
\STATE {\bf CLEAN loop ($t \geq 1$).} \\
While the CLEAN stopping rule is not verified:
\begin{enumerate}
    \item[a.] {\bf{Maximum Identification:}}
        \begin{equation}\label{step-1}
\boldsymbol{x}_{\max}^{(t)} = \arg\max_{\boldsymbol{x}} {\hat{{\boldsymbol{f}}}}^{(t-1)}(\boldsymbol{x}),~~~~~~~~~~~{\hat{{\boldsymbol{f}}}}^{(t)}_{\max} = {\hat{{\boldsymbol{f}}}}^{(t-1)}(\boldsymbol{x}_{\max}^{(t)})~.
\end{equation}
  \item[b.]   {\bf{CLEAN Components Update:}}
\begin{equation}\label{step-2}
{\tilde{{\boldsymbol{f}}}}^{(t)} (\boldsymbol{x}) = {\tilde{{\boldsymbol{f}}}}^{(t-1)}(\boldsymbol{x}) + \frac{\gamma {\hat{{\boldsymbol{f}}}}^{(t)}_{\max}}{\max_{\boldsymbol{x}} |K(\boldsymbol{x})|} \delta(\boldsymbol{x}-\boldsymbol{x}_{\max}^{(t)})~. \hspace{0.1cm} 
\end{equation}
  \item[c.]  {\bf{Dirty Map Update:}} 
\begin{equation}\label{step-3}
{\hat{{\boldsymbol{f}}}}^{(t)}(\boldsymbol{x}) = {\hat{{\boldsymbol{f}}}}^{(t-1)}(\boldsymbol{x}) - \frac{\gamma {\hat{{\boldsymbol{f}}}}^{(t)}_{\max}}{\max_{\boldsymbol{x}} |K(\boldsymbol{x})|} K(\boldsymbol{x}-\boldsymbol{x}_{\max}^{(t)}).
\end{equation}
\end{enumerate}
\STATE {\bf{Estimate of the background map:}}
\begin{equation}\label{background}
{{{\boldsymbol{f}}^B}}(\boldsymbol{x}) \simeq \frac{{\hat{{\boldsymbol{f}}}}(\boldsymbol{x})}{T},
\end{equation}
where ${\hat{{\boldsymbol{f}}}}$ is the result of the final dirty map update and $T$ is an estimate of the integral of the PSF over the Field of View.
\STATE {\bf{Construction of the CLEANed map:}}
\begin{equation}\label{CLEAN-final}
{\overline{{\boldsymbol{f}}}}(\boldsymbol{x}) = (K^C * ({\tilde{{\boldsymbol{f}}}}+{{{\boldsymbol{f}}^B}}))(\boldsymbol{x})=(K^C * {\tilde{{\boldsymbol{f}}}})(\boldsymbol{x}) +  (K^C * {{{\boldsymbol{f}}^B}})(\boldsymbol{x}),
\end{equation}
where ${\tilde{{\boldsymbol{f}}}}$ is the result of the final CLEAN Components update, and  $K^C(\boldsymbol{x})$ is the so-called CLEAN beam, i.e., an idealized version of the PSF.
\end{algorithmic}
\label{alg:alg1}
\end{algorithm}

The automation of the CLEAN loop (Step 2) is guaranteed by a stopping rule that applies when the \lq Dirty Map Update' step returns just experimental noise. Instead, Step 4 of the pipeline, i.e., the construction of the CLEANed map, is clearly ambiguous and mostly biased by the user's decision about the shape of the CLEAN beam $K^C$. In the version of the CLEAN code originally developed for RHESSI \citep{Schmahl}, this convolution kernel is modelled by a two dimensional Gaussian function whose FWHM is chosen by the user according to heuristic rules of thumb. This convolution product is the main reason of the low photometric reliability of the CLEANed map, while conservative choices for FWHM typically lead to under-resolved reconstructions, with correspondingly high $\chi^2$ values. 

In order to solve this issue, u-CLEAN replaces Step 3 and Step 4 of the CLEAN pipeline by a feature augmentation and a soft thresholding step, both characterized by a high degree of automation. 

\section{u-CLEAN via feature augmentation}
\label{uclean}

The u-CLEAN approach is based on interpolating the experimental visibilities with a basis that depends on the CLEAN components map, and on inverting the interpolated visibility surface by means of an iterative constrained algorithm. Specifically, any interpolant, namely $P$, has the property of matching the given measured visibilities at the corresponding locations (the $({\mbox{u}},{\mbox{v}})$ points), i.e.,
\begin{equation}\label{eq1}
P({\boldsymbol{u}}_k)= {\boldsymbol{V}}_k = (\Re({\boldsymbol{V}}_k), \Im({\boldsymbol{V}}_k)), \quad k=1,\ldots,n. 
\end{equation}
In most cases, $P$ consists of a linear combination of $n$ linearly independent basis functions $\{B_1({\boldsymbol{u}}),\ldots,B_n({\boldsymbol{u}})\}$ and can be written as: \begin{equation}\label{eq2}
P({\boldsymbol{u}}) = \sum_{k=1}^{n} (a_k + i~ b_k) B_k({\boldsymbol{u}})~,
\end{equation}
where the coefficients $\{ a_k \}_{k=1}^{n}$ and $\{ b_k \}_{k=1}^{n}$ are determined thanks to the interpolation conditions (\ref{eq1}). Hence, in our case, $P({\boldsymbol{u}})$ approximates the unknown value of the visibility at any given query point ${\boldsymbol{u}}=(u,v)$. By taking a grid of query points, the final output of the interpolation procedure will be an $N \times N$ visibility surface grid, with $N \gg n$. 

\begin{algorithm}[t]
\small
\caption{\textbf{u-CLEAN steps}}
   \hspace*{\algorithmicindent} {\textbf{Inputs:}  Sampling points $ \{ \boldsymbol{u}_k=(u_k,v_k) \}_{k=1}^{n}$; visibility vector $\boldsymbol{V}$; {gain factor} $\gamma$. \\
    \hspace*{\algorithmicindent} \textbf{Outputs:} The u-CLEANED map ${\overline{{\boldsymbol{f}}}}$. }
\begin{algorithmic}[1]
\STATE {\bf{Initialization:}} same as in Algorithm \ref{alg:alg1}.
\STATE {\bf CLEAN loop}: same as in Algorithm \ref{alg:alg1}.
\STATE {\bf{Feature augmentation:}} generate the visibility surface 
\begin{equation}
{\overline{\boldsymbol{V}}} := P(\boldsymbol{u}^{\Psi}) = P((u,v, \Psi(u,v))) = \sum_{k=1}^{n} (a_k + i~ b_k) B_k(u,v, \Psi(u,v)), \quad \Psi \equiv {\boldsymbol{\tilde{V}}} ~.
\end{equation}
\STATE {\bf{Soft thresholding:}} 
     \begin{equation}\label{c1}
    {\overline{{\boldsymbol{f}}}}^{(k+1)} = {\cal{P}}_+[{\overline{\boldsymbol{f}}}^{(k)} + {\overline{\boldsymbol{F}}}^T({\overline{\boldsymbol{V}}} -
    {\overline{\boldsymbol{F}}}{\overline{\boldsymbol{f}}}^{(k)})]~,
    \end{equation}
    where ${\cal{P}}_+$ pixel-wise imposes a positivity constraint, i.e., it returns zero for each negative pixel value and where ${\overline{\boldsymbol{F}}}$ is the $N \times N$ discretized Fourier transform and ${\overline{\boldsymbol{f}}}$ is the $N^2 \times 1$ vector to reconstruct.
\end{algorithmic}
\label{alg:alg2}
\end{algorithm}

In our interpolation procedure we will take advantage of feature augmentation schemes \citep{Bozzini}. Their basic idea consists in the construction of enriched datasets obtained by concatenating the original data with other features that include prior information. In other words, we consider the transformed set of data 
$ \{ \boldsymbol{ u}^{\Psi}_k=(u_k,v_k, \Psi(u_k,v_k)) \}_{k=1}^{n}$, where, for this specific task, $\Psi: \mathbb{C} \rightarrow \mathbb{C}$ depends on the CLEAN components map. Precisely, given the CLEAN components map, namely $\boldsymbol{\tilde{f}}$, which provides a very preliminary knowledge on the flaring source, we define the function $\Psi$ by applying forward to $\boldsymbol{\tilde{f}}$ the Fourier operator in equation (\ref{eq0}), i.e. we get an $N \times N$ grid $\tilde{\boldsymbol{V}}$. Hence, we simply define the scaling function $\Psi$ as the so-computed grid ${\boldsymbol{\tilde{V}}}$. 
Finally, our interpolant will be of the form:
\begin{equation}\label{eq4}
{\overline{\boldsymbol{V}}} := P(\boldsymbol{u}^{\Psi}) = P((u,v, \Psi(u,v))) = \sum_{k=1}^{n} (a_k + i~ b_k) B_k(u,v, \Psi(u,v)), \quad \Psi \equiv {\boldsymbol{\tilde{V}}} ~,
\end{equation}
where our basis $\{B_1({\boldsymbol{u}^{\Psi}}),\ldots,B_n(\boldsymbol{u}^{\Psi})\}$ is given by the so-called Variably Scaled Kernels, e.g. Gaussians or Mat\'ern radial basis functions \citep{bozzini1,vskmpi}. Hence, the u-CLEAN interpolation basis integrates the CLEAN components map via the function $\Psi$. For further details, we refer the reader to the u-CLEAN pipeline which is sketched in  Algorithm \ref{alg:alg2}.

\begin{figure}
    \centering
      \includegraphics[width=12.cm]{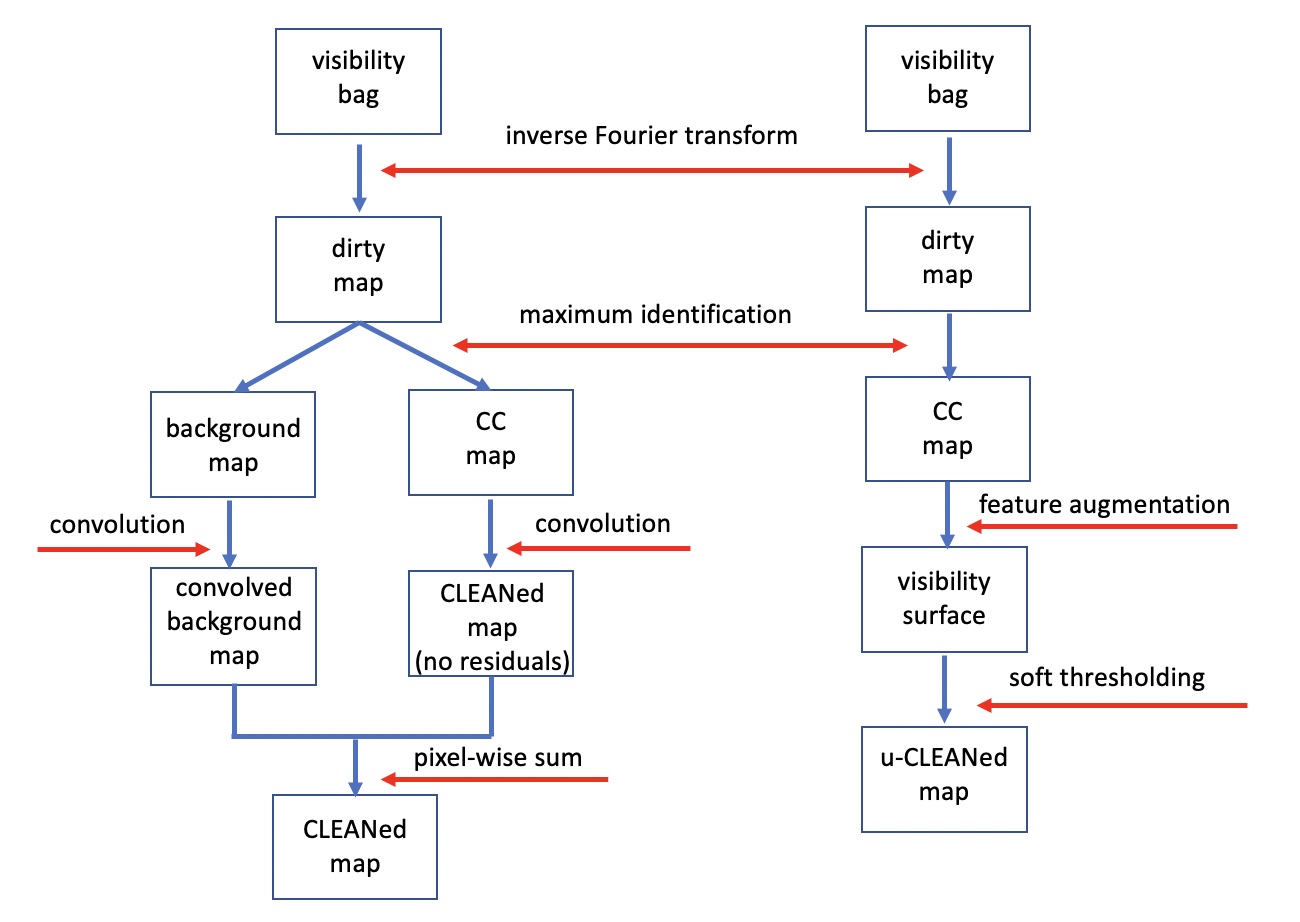}
    \caption{The CLEAN and u-CLEAN pipelines.}
   \label{fig:figure-1}
\end{figure}

Figure  \ref{fig:figure-1} compares the CLEAN and u-CLEAN pipelines, pictorially showing the higher simplicity of the latter one. Further, the new steps in u-CLEAN are completely user-independent: feature augmentation is just an interpolation process, while soft thresholding is a standard projected Landweber scheme \citep{1997InvPr..13..441P} that just needs the choice of an inizialization and the application of a stopping rule. In the current implementation of u-CLEAN we have chosen ${\overline{{\boldsymbol{f}}}}={\boldsymbol{0}}$ and a stopping rule that relies on a check on the $\chi^2$ values \citep{Massone1}.

\section{Application to STIX visibilities}
\label{esperimenti}
\begin{figure}
    \centering
      \includegraphics[scale=0.5]{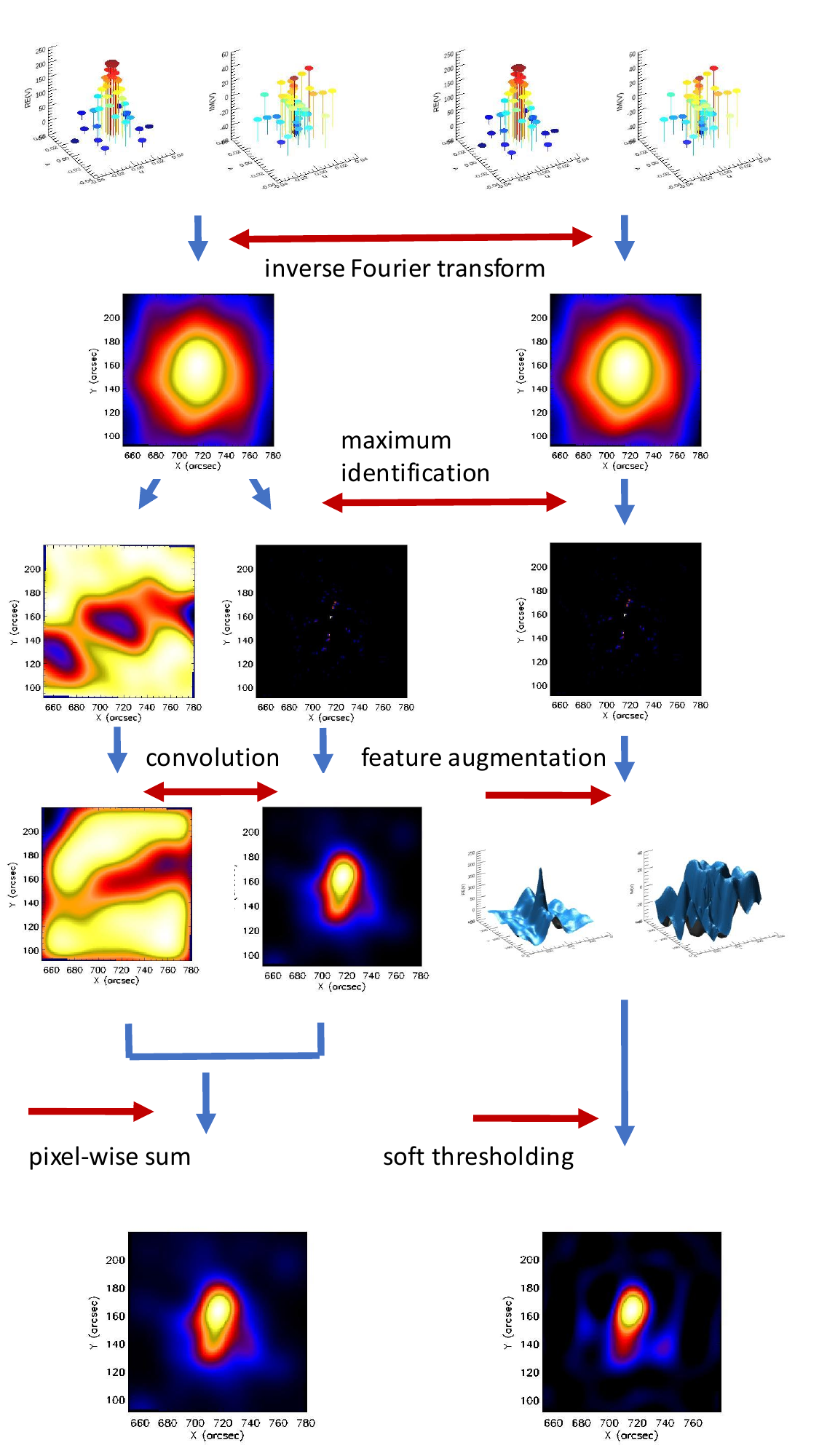}
    \caption{Results of the application of the CLEAN and u-CLEAN pipelines to the STIX visibility bags associated to the November 11 2022 flare in the time range between 01:30:15 and 01:30:45 UT, for the energy channel between 5 and 9 keV.}
   \label{fig:figure-2}
\end{figure}

\begin{figure}
    \centering
      \includegraphics[scale=0.22]{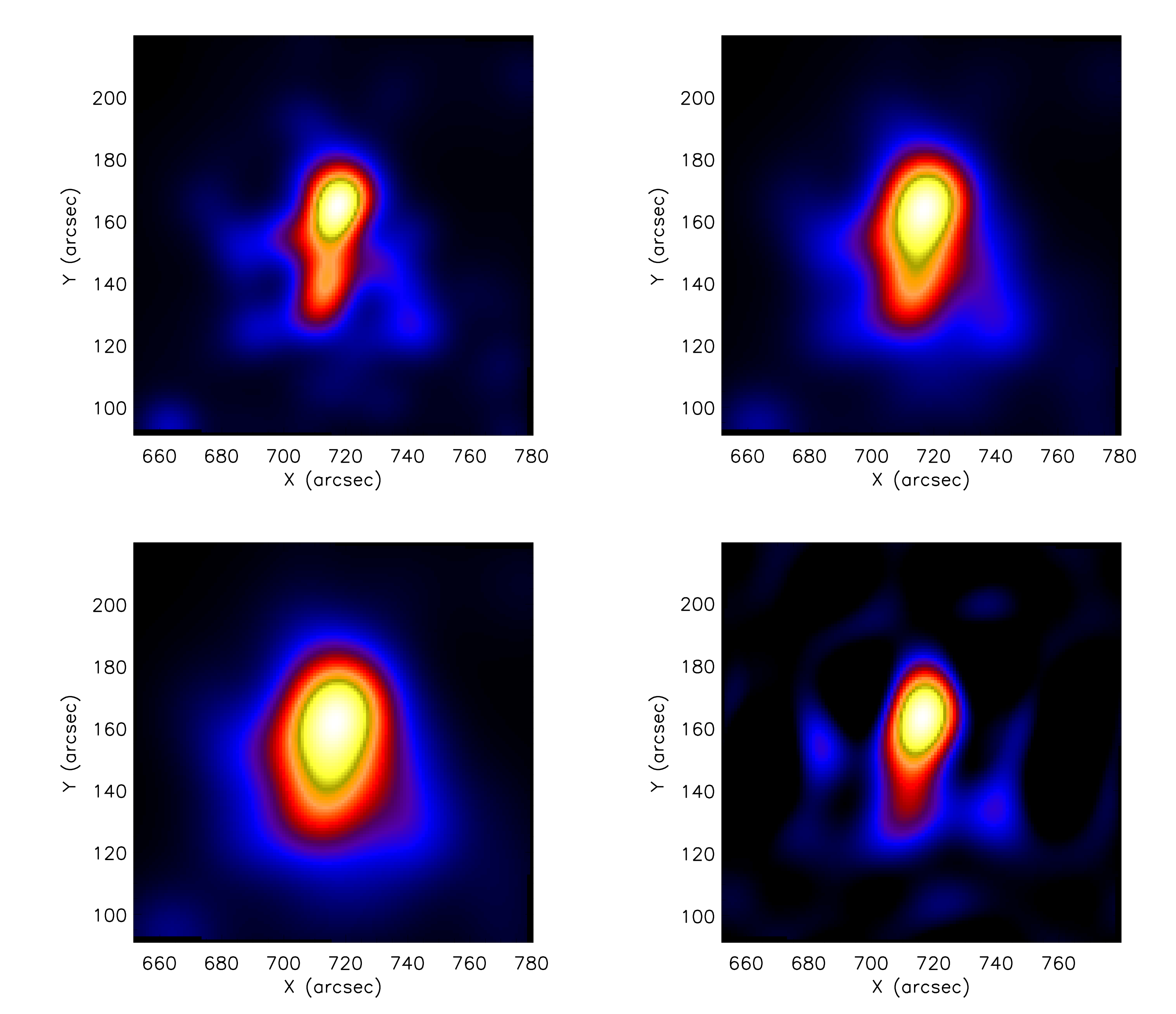}
    \caption{Reconstruction of the November 11 2022 flare in the time range between 01:30:15 and 01:30:45 UT. Top left panel: CLEANed map with FHWM equal to $15$ arcsec; top right panel: CLEANed map with FWHM equal to $20$ arcsec (default value); bottom left panel: CLEANed map with FWHM equal to $25$ arcsec; bottom right panel: u-CLEANed map.}
   \label{fig:novclen}
\end{figure}

\begin{figure}
    \centering
      \includegraphics[scale=0.3]{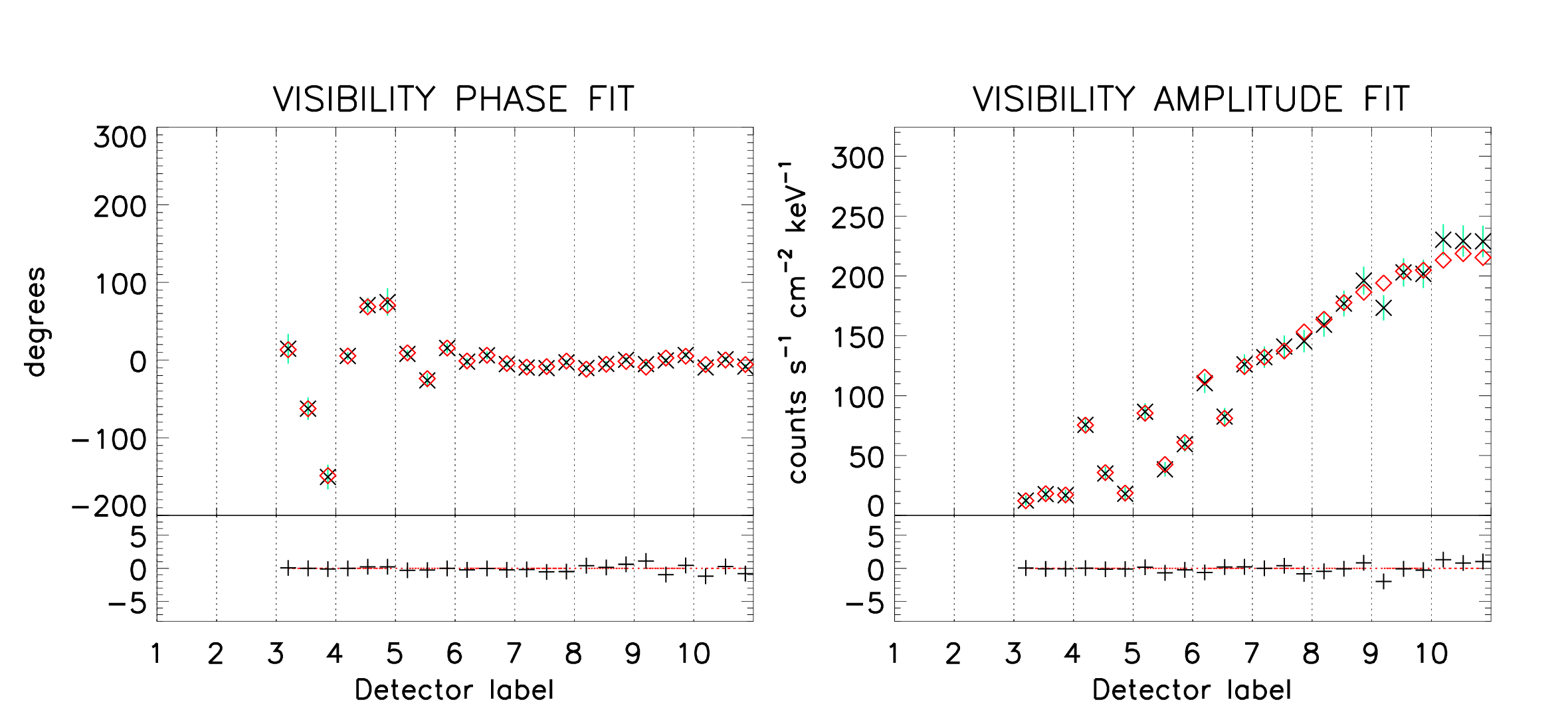}
      \includegraphics[scale=0.3]{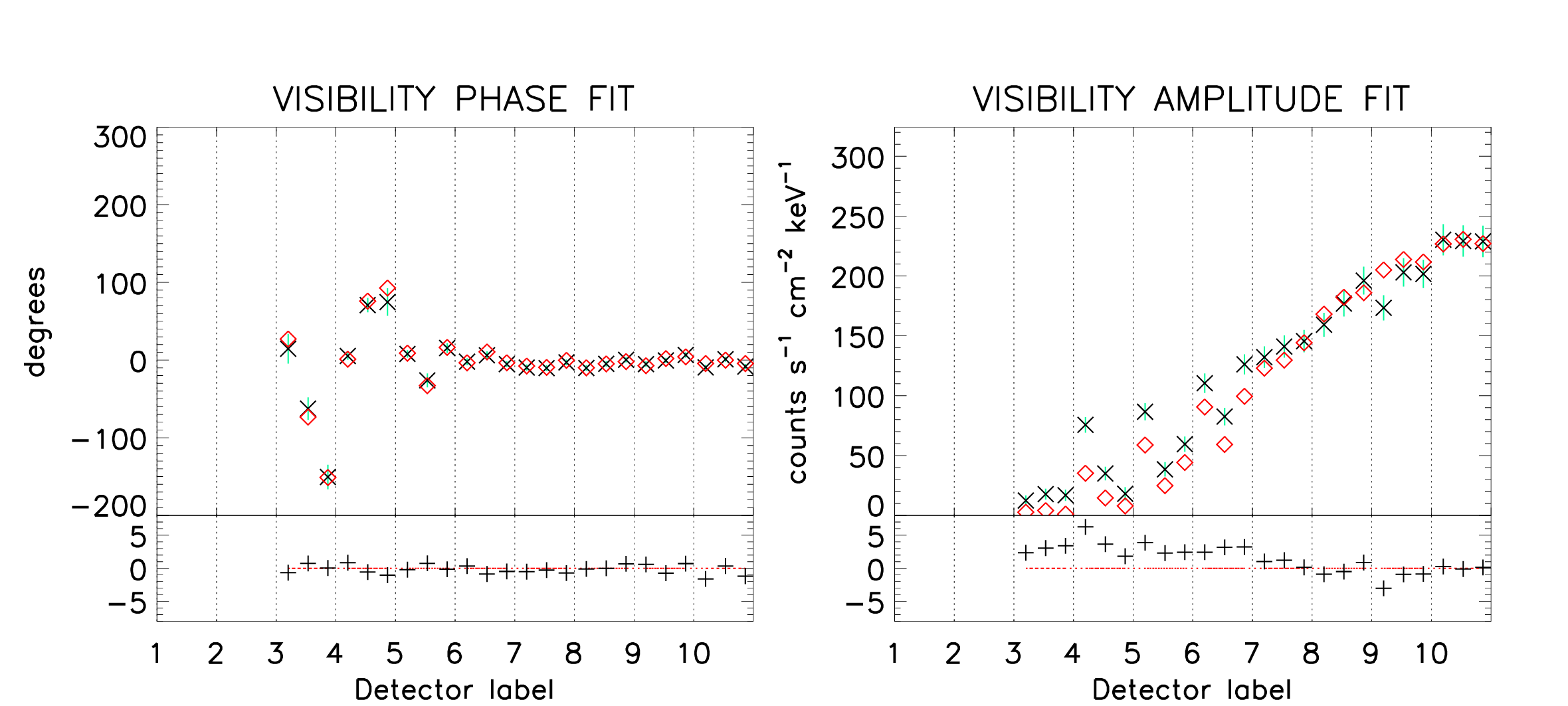}
      \includegraphics[scale=0.3]{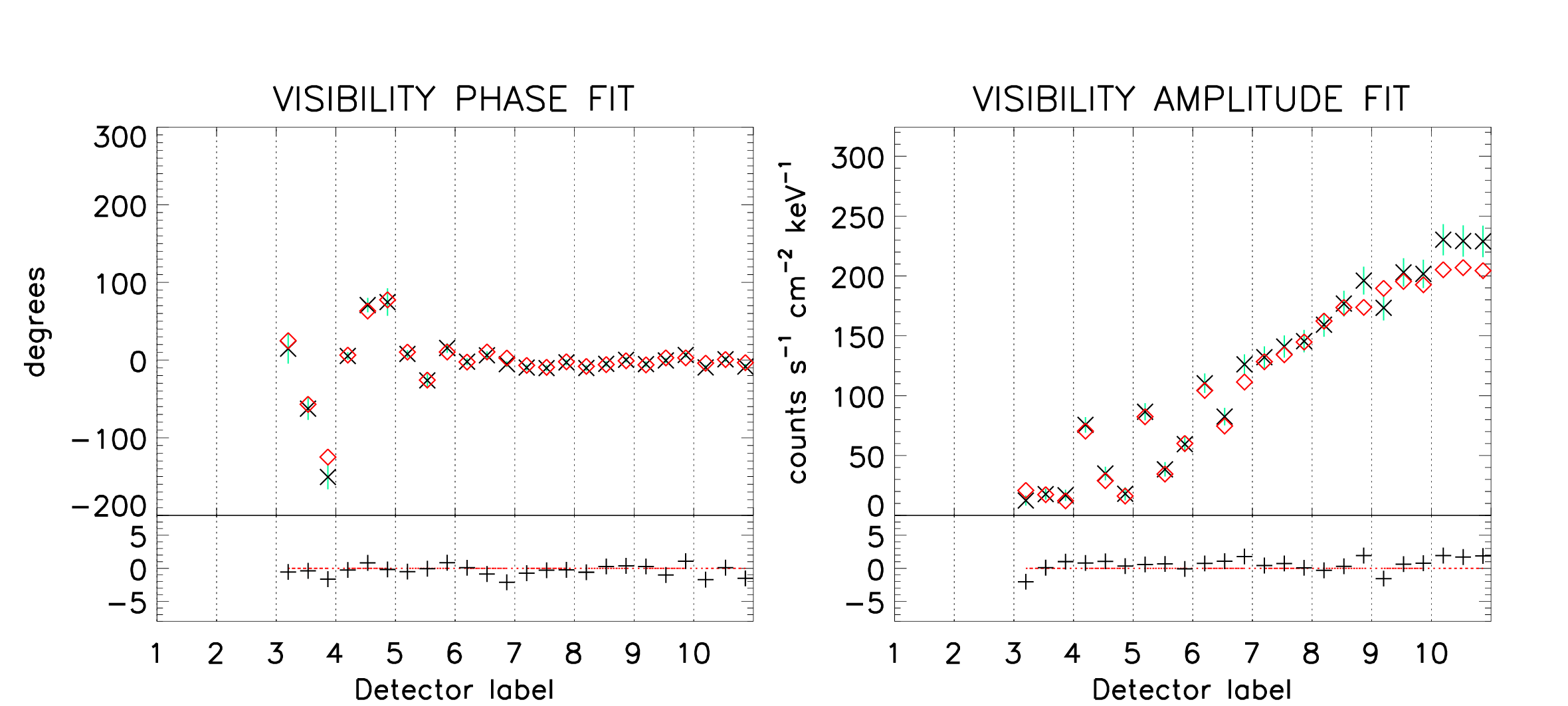}
        \caption{Top to bottom: comparison between the experimental visibilities in the case study of November 11 2022, and the visibilities predicted by the CLEAN components map, and by the images reconstructed with CLEAN and u-CLEAN, respectively.}
   \label{fig:figure-3}
\end{figure}

As a case study we considered the event occurred on November 11 2022 and, at first, we focused on the thermal energy channel 5-9 keV and on the time range from 01:30:15 to 01:30:45 UT. The results of the application of CLEAN and u-CLEAN to the visibility bags associated to this event are illustrated in Figure \ref{fig:figure-2}, which reproduces the scheme of Figure \ref{fig:figure-1} and where, this time, each box contains the actual product of the corresponding computational step. Figure \ref{fig:novclen} compares the reconstruction provided by u-CLEAN with three CLEANed maps obtained by using three different values of the FWHM for the CLEAN beam. Figure \ref{fig:figure-3} contains the fits of the experimental visibilities provided by the CLEAN components map, the CLEANed map (with FWHM for the beam equal to $20$), and the reconstruction provided by u-CLEAN, respectively. 

%\textcolor{magenta}{LA FRASE CHE SEGUE LA TOGLIEREI PERCHE LA SOSTITUIREI CON TABELLA \ref{tab:novemberclean}.}
%The corresponding overall $\chi^2$ values are $0.71$ for the components map (\textcolor{magenta}{denoted by CC in what follows), $7.09$ for the CLEANed map, and $2.47$ for the u-CLEAN reconstruction (usually, in the IDL code stx$\_$vis$\_$clean implementing CLEAN, the $\chi^2$ values correspond to the CLEAN components map).} 
We point out that, in the CLEAN pipeline of Figure \ref{fig:figure-2}, the residual map (left panel after \lq maximum identification') corresponds to the final iteration of the \lq Dirty Map Update'. We also notice that the two panels after \lq feature augmentation' in the u-CLEAN pipeline correspond to the real and imaginary parts of the interpolated visibility surface. Further, Figure \ref{fig:novclen} and Table \ref{tab:novemberclean} show that the CLEANed and u-CLEAN reconstructions are characterized by a similar overall shape and photometric values of the same order of magnitude. However, the u-CLEAN reconstruction is better resolved, while, in the case of the CLEANed maps, the resolution and the corresponding $\chi^2$ values get worse for increasing FWHM values in $K^C$. We point put that we also report the $\chi^2$ values computed with the Clean Component (CC) map, as usually the community relies on that values.

\begin{table}[]
    \centering
    \begin{tabular}{c|c|c|c|c|c}
    & CCs & CLEAN & CLEAN  & CLEAN  &  u-CLEAN \\
    &  &  FWHM $15$  & FWHM $20$ & FWHM $25$ &  \\
    \hline\hline
    November 11 (5-9 keV) & $0.70$ & $5.03$ & $6.92$ & $11.04$ & $2.08$   \\
    \hline
    \hline
    \end{tabular}
    \caption{$\chi^2$ values for the reconstructions of the November 11 2022 flare in the time range between 01:30:15 and 01:30:45 UT, provided by the CLEAN components map (CCs), by CLEAN (for three different values of FWHM measured in arcsec) and by u-CLEAN.}
    \label{tab:novemberclean}
\end{table}

The second experiment compared u-CLEAN performances to the ones provided by other reconstruction methods in the case of some flaring events observed by STIX in both the thermal and non-thermal regimes. Specifically, we considered
\begin{itemize}
    \item The May 8 2021 event in the time range between 18:24:00 and 18:32:00 UT, in the thermal channel between 6 and 10 keV.
    \item The same event as before in the same time interval but, this time, in the non-thermal channel between 18 and 28 keV.
    \item The June 7 2020 event in the time range between 21:39:00 and 21:42:49 UT and in the thermal range between 6 and 10 keV.
    \item The March 31 2022 event in the time interval between 18:26:20 and 18:27:00 UT and in the non-thermal energy range between 25 and 50 keV.
\end{itemize}

\begin{figure}
    \centering
      \includegraphics[scale = 0.25]{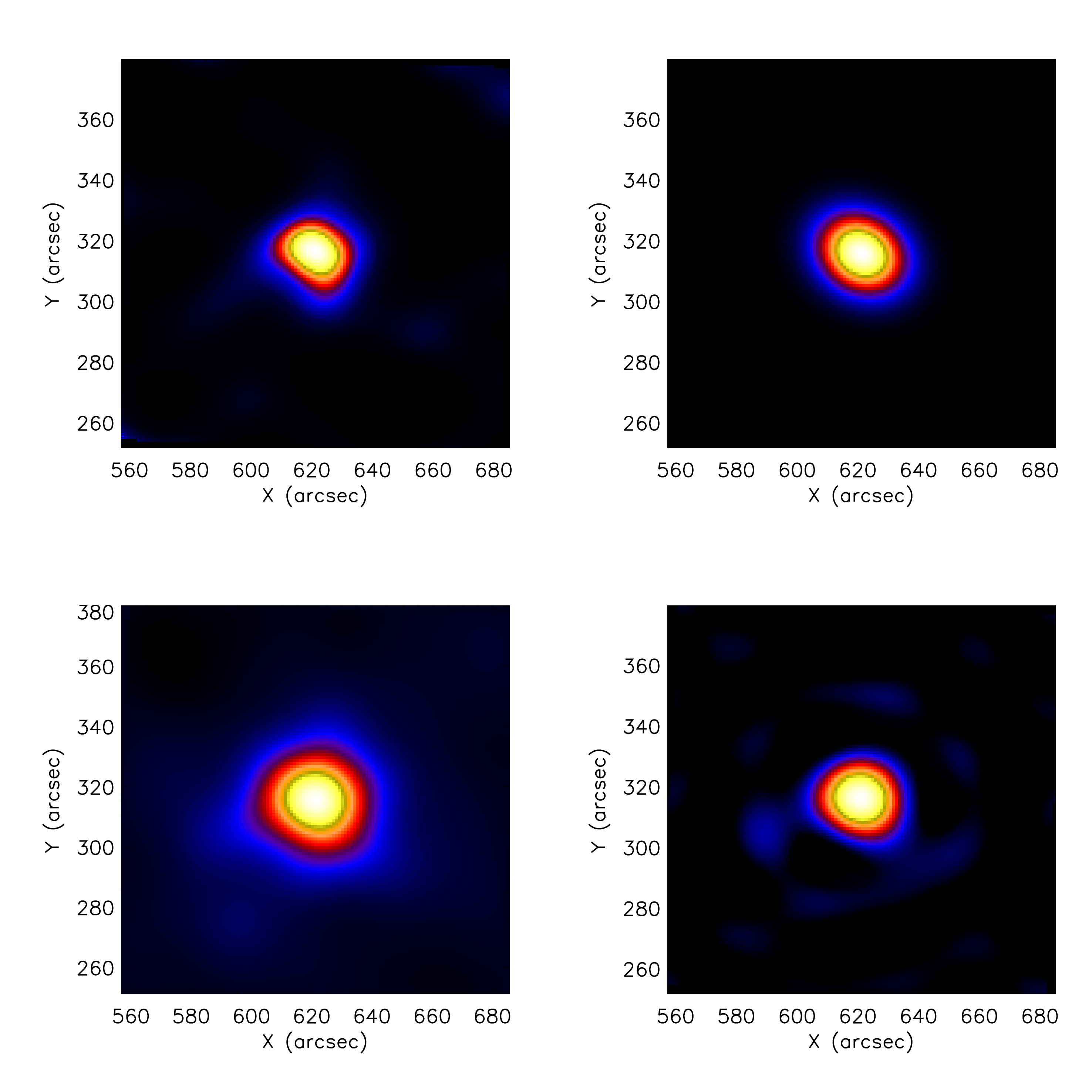}
    \caption{Reconstructions of the the May 8 2021 event in the energy range 5-9 keV provided by MEM$\_$GE (top left panel), VIS$\_$FWDFIT$\_$PSO (top right panel), CLEAN (bottom left panel), and u-CLEAN (bottom right panel).}
   \label{fig:thermal}
\end{figure}

\begin{figure}
    \centering
      \includegraphics[scale = 0.25]{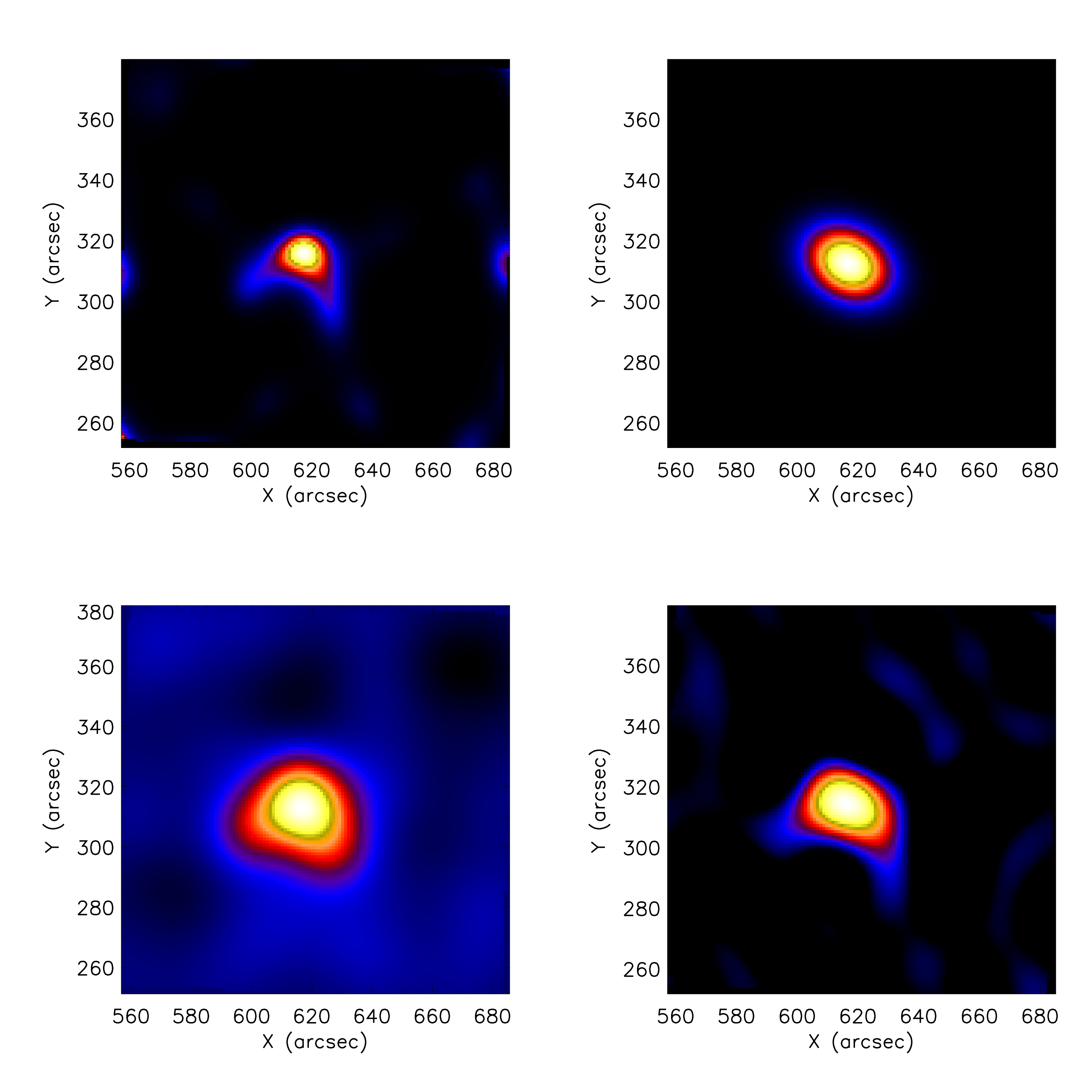}
    \caption{Reconstructions of the May 8 2021 event in the energy range 18-28 keV provided by MEM$\_$GE (top left panel), VIS$\_$FWDFIT$\_$PSO (top right panel), CLEAN (bottom left), and u-CLEAN (bottom right).}
   \label{fig:non-thermal}
\end{figure}

\begin{figure}
    \centering
      \includegraphics[scale = 0.25]{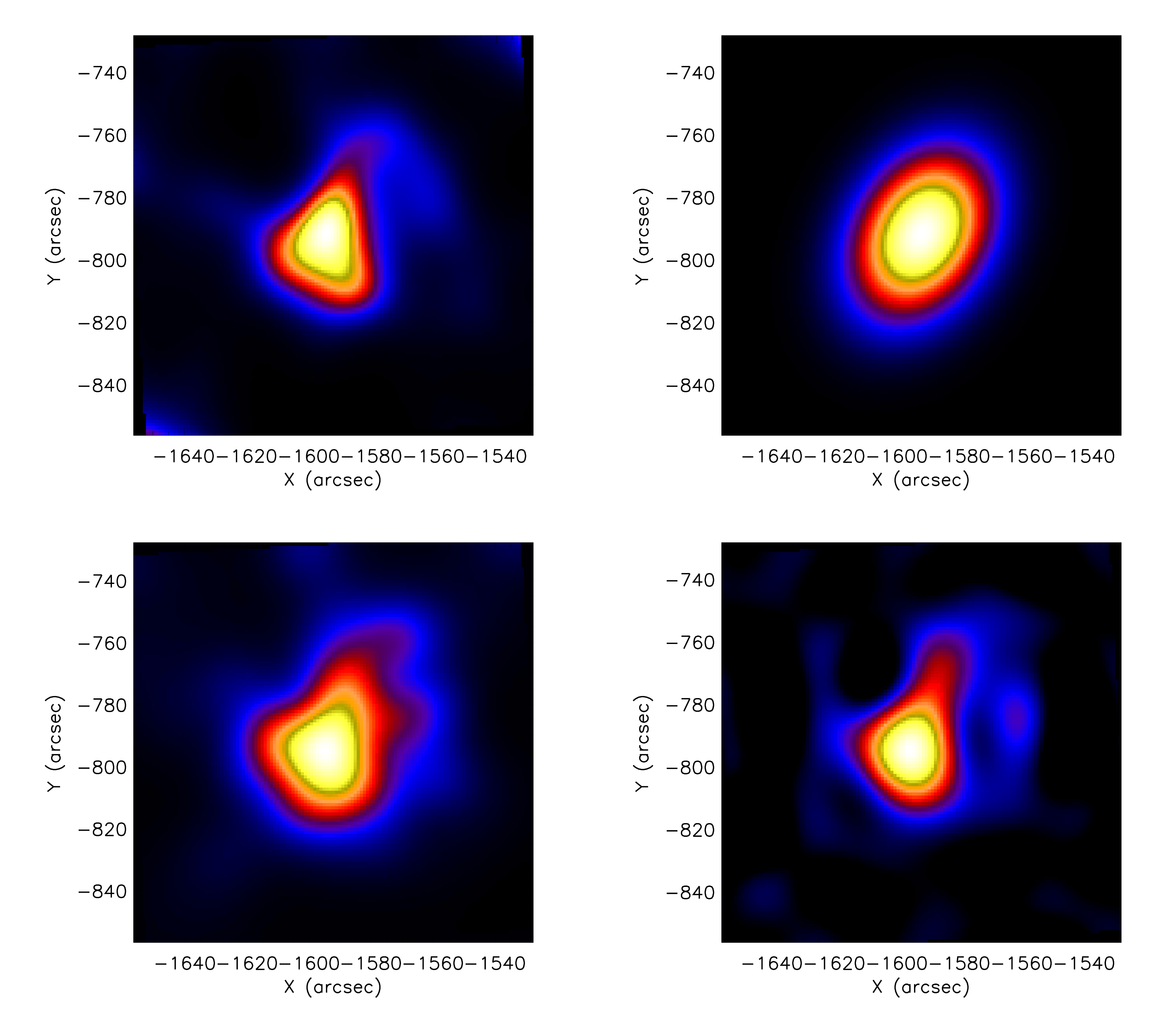}
    \caption{Reconstructions of the June 7 2020 event in the energy range 6-10 keV provided by MEM$\_$GE (top left), VIS$\_$FWDFIT$\_$PSO (top right), CLEAN (bottom left), and u-CLEAN (bottom right).}
   \label{fig:june7}
\end{figure}

\begin{figure}
    \centering
      \includegraphics[scale = 0.25]{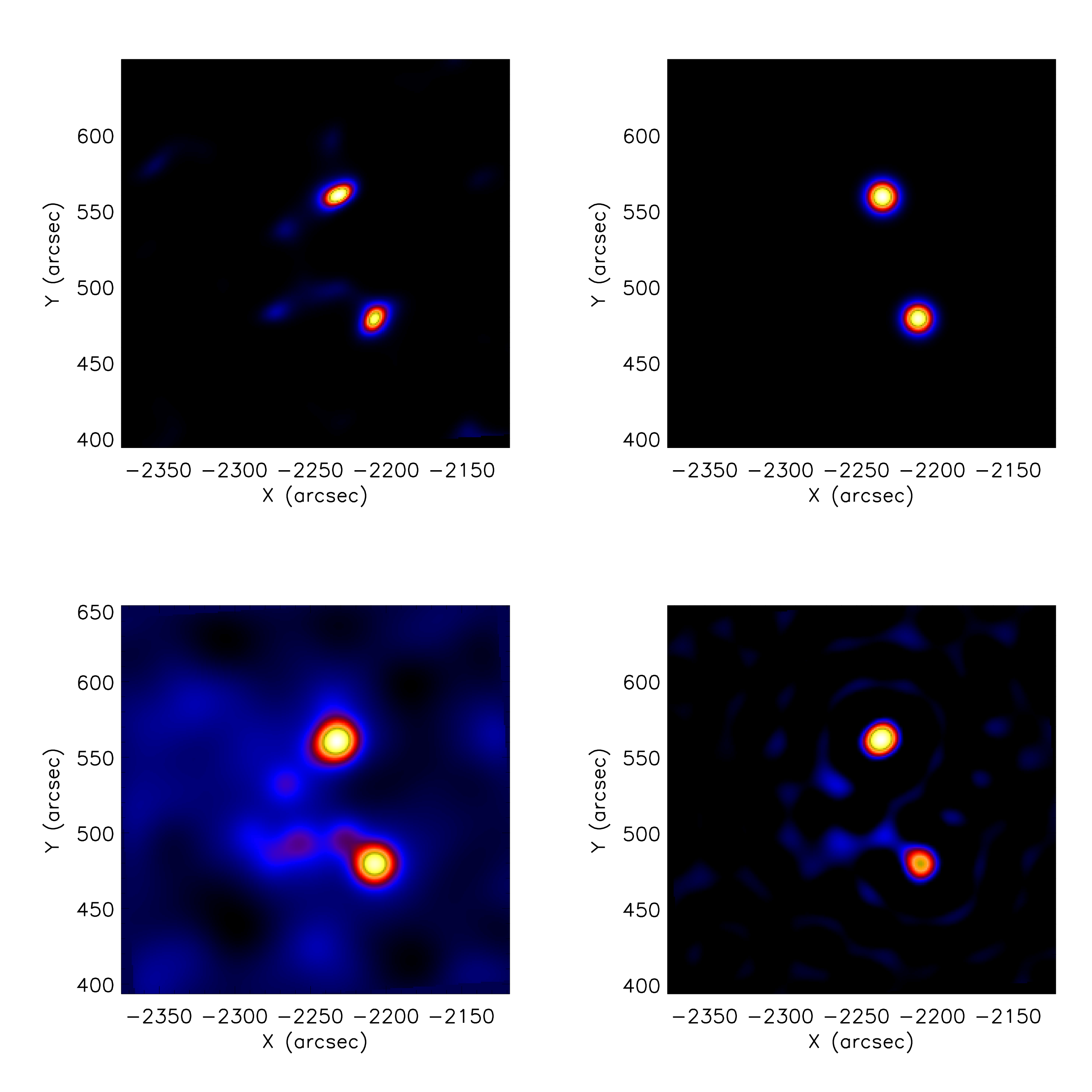}
    \caption{Reconstructions of the March 31 2022 event in the energy range 25-50 keV provided by MEM$\_$GE (top left), VIS$\_$FWDFIT$\_$PSO (top right), CLEAN (bottom left), and u-CLEAN (bottom right).}
   \label{fig:march31}
\end{figure}

The reconstructions of these flaring sources are represented in Figure \ref{fig:thermal}, Figure \ref{fig:non-thermal}, Figure \ref{fig:june7}, and Figure \ref{fig:march31}, and have been obtained by means of MEM$\_$GE \citep{massa2020mem_ge}, VIS$\_$FWDFIT$\_$PSO \citep{2022A&A...668A.145V}, CLEAN, and u-CLEAN, respectively. The corresponding $\chi^2$ values are contained in Table \ref{tab:table-3}. We point out that, in the thermal cases, VIS$\_$FWDFIT$\_$PSO utilizes an elliptical Gaussian function as input model, and that, for all cases, FWHM in CLEAN is set to the default value of $20$ arcsec. These experiments show that MEM$\_$GE and u-CLEAN provide similar results in terms of both fitting reliability and overall morphology. In particular, in Figure \ref{fig:non-thermal} the reconstructions provided by the two methods nicely follow the loop shape of the source. Further, the foot-points in Figure \ref{fig:march31} are probably over-resolved by MEM$\_$GE and under-resolved by CLEAN, while VIS$\_$FWDFIT$\_$PSO and u-CLEAN provide similar results.

\begin{table}[]
    \centering
    \begin{tabular}{c|c|c|c|c|c}
    & MEM$\_$GE & VIS$\_$FWDFIT$\_$PSO & CCs & CLEAN & u-CLEAN \\
    \hline\hline
    May 8 (5-9 keV) & $2.75$ & $5.66$ & $3.97$ & $32.5$ & $4.75$ \\
    \hline
    May 8 (18-28 keV) & $2.96$ & $3.29$ & $2.84$ & $4.04$ & $3.19$ \\
    \hline
    June 7 (6-10 keV) & $1.73$ & $4.38$ & $0.99$ & $3.22$ & $1.72$ \\
    \hline
    March 31 (25-50 keV) & $3.99$ & $5.26$ & $1.96$ & $11.6$ & $6.17$ \\
    \hline
    \end{tabular}
    \caption{$\chi^2$ values predicted by MEM$\_$GE, VIS$\_$FWDFIT$\_$PSO, the CLEAN components map, the CLEANed map (with FWHM equal to the deafault value of $20$ arcsec) and by u-CLEAN in the cases of the May 8 2021, June 7 2020, and March 31 events. For the May 8 event both the thermal channel between $5$ and $10$ keV and the non-thermal one between $18$ and $28$ keV are considered.}
    \label{tab:table-3}
\end{table}

\section{Comments and conclusions}
\label{conclusioni}
Although CLEAN is a reference tool for image reconstruction in hard X-ray solar physics, its reliability and automation degree are significantly limited by the final convolution step involving the CLEAN components map and an idealized model for the instrument PSF, and by the need to a posteriori add a background estimate to the result of this deconvolution. These two rather heuristic steps may imply under-resolved CLEANed maps, with unrealistically high $\chi^2$ values, and an unsatisfactory automation level for the overall algorithm. Therefore, the present study exploits the CLEAN components map to generate an interpolated visibility surface, and applies soft thresholding to reconstruct an unbiased CLEAN map in the image space. The resulting u-CLEAN is an iterative scheme characterized by a high degree of automation and by reconstruction performances in line with respect to the ones provided by other imaging methods developed for STIX. u-CLEAN is now at disposal for testing at the URL \url{https://github.com/theMIDAgroup/U-CLEAN} of the STIX ground software. Further, we believe that a multi-scale extension is of rather straightforward implementation. Finally, we point out that u-CLEAN can be interpreted as the first release of the feature augmented uv$\_$smooth tailored to the case of STIX visibilities. 

\begin{acknowledgements}
{\em{Solar Orbiter}} is a space mission of international collaboration between ESA and NASA, operated by ESA. The STIX instrument is an international collaboration between Switzerland, Poland, France, Czech Republic, Germany, Austria, Ireland, and Italy. AMM, AV, and MP are supported by the \lq Accordo ASI/INAF Solar Orbiter: Supporto scientifico per la realizzazione degli strumenti Metis, SWA/DPU e STIX nelle Fasi D-E'. EP and AMM acknowledge the support of the Fondazione Compagnia di San Paolo within the framework of the Artificial Intelligence Call for Proposals, AIxtreme project (ID Rol: 71708). AMM is also grateful to the HORIZON Europe ARCAFF Project, Grant No. 101082164.
 
% SG acknowledges the financial support from the ``Accordo ASI/INAF Solar Orbiter: Supporto scientifico per la realizzazione degli strumenti Metis, SWA/DPU e STIX nelle Fasi D-E''.
\end{acknowledgements}

\bibliographystyle{aa}
\bibliography{bib_stix}

\end{document}